\documentclass[preprint2]{aastex6}

\usepackage{graphics}
\usepackage{natbib}
\usepackage{amsmath}
\newcommand{\ba}{\begin{eqnarray}}
\newcommand{\ea}{\end{eqnarray}}

\shortauthors{Tamayo et al.}

\begin{document}

\title{Convergent Migration Renders TRAPPIST-1 Long-lived}

\author{Daniel Tamayo\altaffilmark{1,2,3}, Hanno Rein\altaffilmark{1,4},Cristobal Petrovich\altaffilmark{2,5}, and Norman Murray\altaffilmark{2}}
\altaffiltext{1}{Department of Physical \& Environmental Sciences, University of Toronto at Scarborough, Toronto, Ontario M1C 1A4, Canada}
\altaffiltext{2}{Canadian Institute for Theoretical Astrophysics, 60 St. George St, University of Toronto, Toronto, Ontario M5S 3H8, Canada}
\altaffiltext{3}{Centre for Planetary Sciences Fellow}
\altaffiltext{4}{Department of Astronomy \& Astrophysics, University of Toronto, Toronto, Ontario M5S 3H4, Canada}
\altaffiltext{5}{CITA Fellow}
\email{d.tamayo@utoronto.ca}

\begin{abstract}
TRAPPIST-1 is a late M-dwarf orbited by seven Earth-sized planets with orbital period ratios near a chain of mean motion resonances. Due to uncertain system parameters, most orbital configurations drawn from the inferred posterior distribution are unstable on short timescales of $\sim$ 0.5 Myr, even when including the eccentricity damping effect of tides. By contrast, we show that most physically plausible resonant configurations generated through disk migration are stable even without tidal dissipation on timescales of at least 50 Myr ($10^{10}$ orbits), an increase of at least two orders of magnitude. This result, together with the remarkable chain of period ratios in the system, provide strong evidence for convergent migration naturally emplacing the system near an equilibrium configuration forced by the resonant chain. We provide an openly available database of physically plausible initial conditions for TRAPPIST-1 generated through parametrized planet-disk interactions, as well as bit-by-bit reproducible N-body integrations over $10^9-10^{10}$ orbits.
\end{abstract}


\section{Introduction} \label{intro}
TRAPPIST-1 is a late M-dwarf of mass $M_\star\approx0.08M_\odot$ hosting seven transiting Earth-mass planets in the longest resonant chain known to date, with adjacent planets near period ratios of 8:5, 5:3, 3:2, 3:2, 4:3 and 3:2 \citep{Gillon16, Gillon17, Luger17}. The system is of distinctive interest given that several of the planets are likely in the star's habitable zone and that the star's proximity to us (12 pc) makes the planets amenable to atmospheric characterization with the Hubble Space Telescope \citep{deWit16} and the upcoming James Webb Space Telescope \citep{Barstow16}.

An important tension is that while the star is likely 3-8 Gyr old \citep{Luger17}, \cite{Gillon17} found that N-body integrations with initial conditions drawn from their orbital fits typically went unstable on a timescale of 0.5 Myr, ten thousand times shorter than the system's age. Given the low probability of catching the system in such a transient state and the planets' proximity to the star, \cite{Gillon17} suggested that eccentricity damping due to equilibrium tides raised on the planets could help stabilize the system.

The corresponding eccentricity damping timescale $\tau_e$ is given by \citep[e.g.,][]{Murray99}
\begin{equation}
\tau_e = \frac{4}{63} \frac{\tilde{\mu}Q}{2\pi}\Bigg(\frac{m}{M_\star}\Bigg)\Bigg(\frac{a}{R}\Bigg)^5P,
\end{equation}
where $m$ and $M_\star$ are the masses of the planet and star respectively, $a$ is the planet's semimajor axis, $R$ its physical radius, and $P$ its orbital period; drawing values for these quantities from \cite{Gillon17}, and assuming an effective rigidity $\tilde{\mu} = (10^4\text{km}/R)^2$ and tidal quality factor $Q=100$ for the planet, we find that $\tau_e \approx 0.2 Myr$ for the innermost planet.

While this damping on the innermost planet can be effectively communicated to the outer bodies through their mutual gravitational coupling \citep[e.g.,][]{Hansen15}, $\tau_e$ is comparable to the instability timescale in the absence of dissipation, so it is not clear whether the tidal damping is strong enough to hold the system together. Indeed, \cite{Gillon17} found that in most cases, adding tidal effects to their simulations only delayed the instability. 

In this letter we instead study the role of accurate initial conditions. A major challenge is the high-dimensional parameter space of planetary masses and initial conditions, several of which have large uncertainties. The near-integer period ratios between orbits place the system in a very rich region of phase space where small changes in initial conditions can lead to very different dynamical behavior. Indeed, \cite{Obertas17} recently integrated a large suite of closely packed three-planet systems. They found that near integer period ratios corresponding to strong mean motion resonances (MMRs), initially circular orbits (i.e., far from their resonantly forced eccentricities) the instability time can drop by several orders of magnitude. By contrast, systems initialized near the centers of their respective MMRs in phase space can have their stability significantly enhanced \citep[e.g., the Hilda family of asteroids in the 3:2 MMR with Jupiter][]{Broz08}.

The fact that TRAPPIST-1 is observed today with its near-integer period ratios after $\sim 10^{12}$ dynamical times suggests that the system lies on a long-lived resonant or near-resonant trajectory. However, the simulations of \cite{Gillon17} imply that the fraction of phase space occupied by such trajectories is small, and that the large uncertainties in the planetary masses, eccentricities and apsidal alignments render it unlikely for initial conditions drawn from the high-dimensional inferred posteriors to be long-term stable. 

We argue that this relative scarcity of stable trajectories provides strong evidence for a slow migration history in the TRAPPIST-1 system in order to emplace the planets near the much longer-lived equilibrium centers of their respective resonances. Rather than working from modeled posteriors of the system configuration as observed today, we generate a suite of physically plausible initial conditions through parametrized simulations of disk migration, and investigate their long-term stability. This is important in order to obtain self-consistent initial conditions following the assembly of the resonant configuration, since tidal eccentricity damping will cause the resonant chain to slowly spread apart \citep{Greenberg81, Batygin12, Lithwick12}---the configuration observed today therefore does not provide appropriate initial conditions for integrating the history of TRAPPIST-1.

\section{Resonant Chains From Disk Migration} \label{genic}

If migrating toward one another slowly enough \citep{Quillen06, Mustill11, Ogihara13}, and with low enough eccentricities \citep{Henrard82, Borderies84, Batygin15}, a pair of planets experiencing eccentricity damping from a protoplanetary disk will capture into MMRs and reach equilibrium eccentricities where the convergent migration torques are balanced by divergent torques due to eccentricity damping \citep{Goldreich14, Deck15overstability, Delisle15}. Such analyses have been extended for multiplanet systems \citep{Papaloizou16}, but potentially depend on many uncertain parameters.

Thus, starting from widely separated orbits and obtaining the observed seven-planet resonant chain from a single set of disk and planet properties would require substantial fine tuning. 
Instead, we developed a simple, easily automatable procedure for migrating the planets into a range of physically plausible initial configurations.

We ran N-body simulations with all planet-pair period ratios initialized $2\%$ wide of their observed commensurabilities on circular orbits. The masses and inclinations were drawn from independent normal distributions based on the values reported by \cite{Gillon17}, the longitudes of ascending node and true longitude were drawn from a uniform distribution.

We then applied an inward migration force (a drag force following the prescription in \citealt{Papa00}) only to the outermost planet on a timescale $\tau_a$ of $6\times10^5$ orbital periods, and eccentricity damping forces (at constant angular momentum, again following \citealt{Papa00}) on a timescale $\tau_e = \tau_a/K$ to all planets, where $K$ \citep{Lee02} was randomly drawn for each system (see Sec.\:\ref{simulations})\footnote{We note that even planets feeling only such eccentricity damping forces will nevertheless migrate (slowly) on a timescale $\tau_e/e^2$. Where ambiguous, we therefore refer to the main migration forces (on timescale $\tau_a$) as ``direct migration forces."} The migration slows down as planets are captured into the resonant chain. This yields effective $K$ values that vary with time, but the qualitative behavior remains clear. For high eccentricity damping strength $K$, there is a powerful resonant repulsion effect \citep{Lithwick12, Batygin12}, so converging orbits stall early at wider period ratios from resonance, and lower equilibrium eccentricities. Conversely, for weak eccentricity damping (low K), planets can migrate deeper into resonance before reaching equilibrium, yielding period ratios closer to the nominal resonant period ratio and higher equilibrium eccentricities. 

The above $\tau_a$ was found by increasing it manually until the migration was slow enough to capture planets into the observed chain of 8:5, 5:3, 3:2, 3:2, 4:3 and 3:2 period ratios a substantial ($\approx 40\%$) fraction of the time. This timescale of $\approx 3\times10^4$ years is much shorter than the typical Myr lifetimes of disks around low-mass stars and brown dwarfs \citep{Jayawardhana03}. We note that if the planets' absolute migration times in the TRAPPIST-1 disk were comparable to the disk timescale, one requires a stalling mechanism to avoid the planets falling into the star.

We integrated for one semimajor axis timescale $\tau_a$, and then adiabatically removed all the damping forces. We then exploited the scale-free nature of Newtonian gravity to rescale the semimajor axes to the observed values.

\begin{figure*}
 \centering \resizebox{\textwidth}{!}{\includegraphics{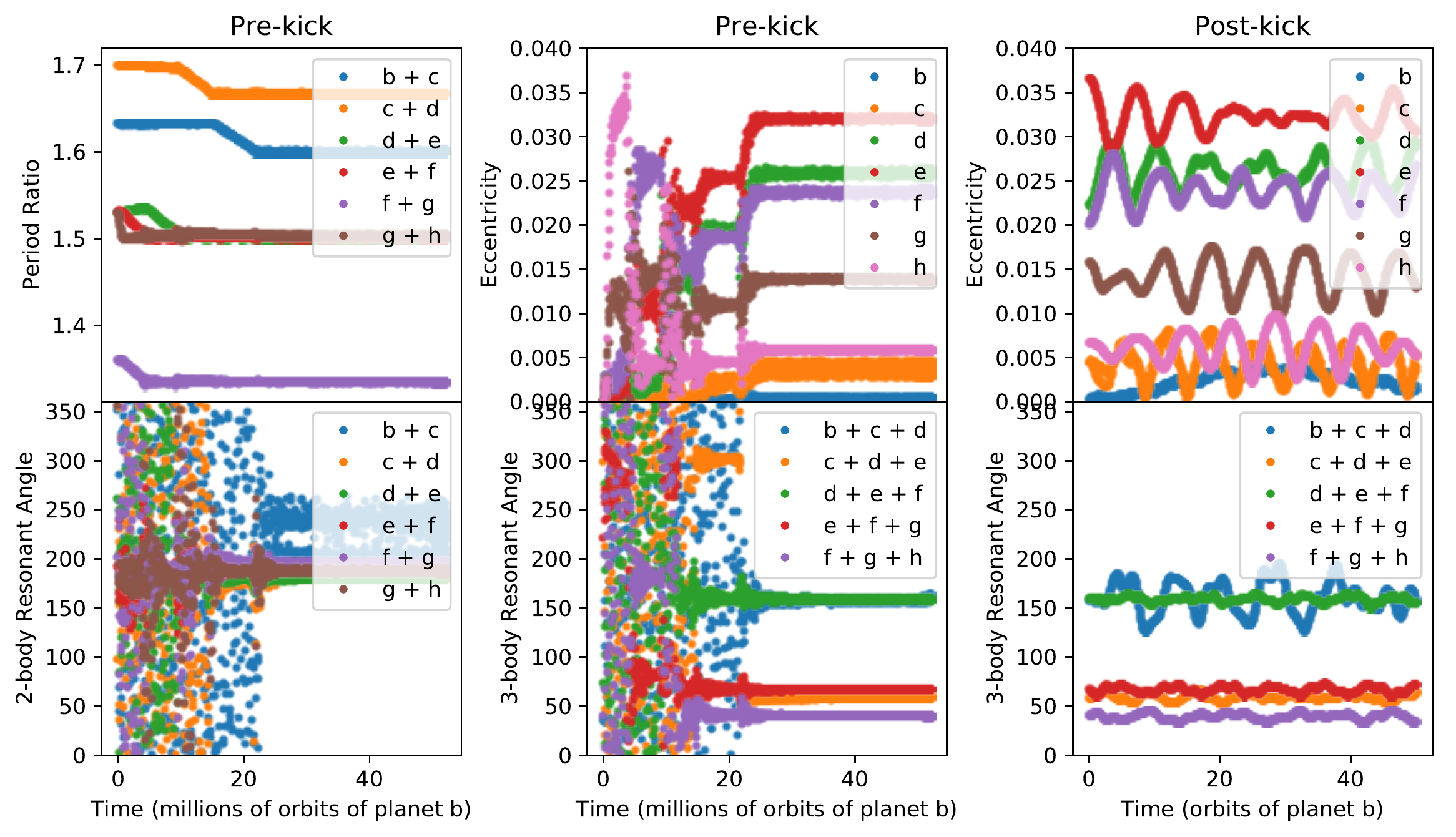}}
 \caption{
     Top left panel plots the period ratios between adjacent pairs of planets vs. time, and shows capture near the observed commensurabilities for each pair of planets. The left and middle panels in the bottom row show the evolution of one of the two-body resonance angles and the three-body resonance angles, respectively (see text). All angles capture into libration with small libration amplitudes. The top middle panel plots the planetary eccentricities, which evolve with each resonance capture. As described in the text, the panels in the rightmost column plot the same quantities as the middle column, but after applying a parametrized turbulent kick (note that the time axis is shorter by a factor of a million). The post-kick panels represent the initial conditions for this particular realization that would be used for the long-term integrations.
\label{migration}}
\end{figure*}

A typical result of the above procedure is shown in Fig.\:\ref{migration}, and illustrates the diversity of possible outcomes. Only planet $h$\footnote{the planets are labeled by successive letters starting from $b$ for the innermost planet} feels a direct migration force, and as can be seen in the period ratio panel, quickly captures into a 3:2 MMR with $g$ (brown line). From that point, the pair migrates inward together more slowly, decreasing the period ratio between $f$ and $g$ (purple) until the pair is captured into a 4:3 MMR. During this period, planets $d$ and $e$ feel no direct migration forces; however, due to the eccentricity damping forces and the associated resonant repulsion \citep{Batygin12, Lithwick12}, they diverge from one another, which pushes $e$ and $f$ into their 3:2 MMR (red). The resonant convoy then picks up $d$, $c$ and $b$ into their respective observed commensurabilities. As seen in the top middle panel, each of these resonant captures tends to raise eccentricities to new equilibrium values, but it also disturbs the existing configuration and can cause eccentricities to decrease (e.g., planet $h$, pink) as the system establishes a new equilibrium. Thus, not only does increasing $K$ lower the equilibrium eccentricities as discussed above, it can also change the order in which resonances are established, leading to a diversity of initial configurations that distribute eccentricities differently among the planets.

The bottom left panel shows one of the 2-body MMR resonant angles, specifically the one of the form $(p+q)\lambda_2 - p\lambda_1 - q\varpi_1$ for the values of $p$ and $q$ appropriate for their respective $p+q:p$ MMR, where $\lambda$ is the mean longitude, $\varpi$ is the longitude of pericenter, and the subscripts refer to the inner and outer planet, respectively. The bottom middle panel plots the nearby three-body resonance angle $p\lambda_1 - (p+q)\lambda_2 + q\lambda_3$ for the values of $p$ and $q$ listed in Table 1 of \cite{Luger17}. We note that the various three-body resonances have different libration centers depending on the parameters drawn to generate the initial conditions.

The planets thus park into their observed commensurabilities with low libration amplitudes, which presents the best-case scenario for stability. As a last step, we qualitatively accounted for the fact that stochastic turbulent perturbations likely yield deviations from exact equilibrium \citep{Rein09, Batygin17}. We drew intuition from the restricted three-body problem, where small offsets (free eccentricities) from the equilibrium eccentricity forced by the MMR yield librations around the resonant center. The separatrix is encountered at free eccentricities comparable to the forced eccentricity. In a resonant chain there are multiple MMRs, but for simplicity, we randomly drew a single parameter $\eta$ from the interval $[0,1]$, and for each planet randomly increased or decreased its total eccentricity by the fraction $\eta$. The evolution following a kick of $\eta \approx 0.13$ (over a much shorter timescale, comparable to the 79 days of K2 observations) can be seen in the two right panels. 

This procedure therefore generated a simple, reproducible and easily automated two-parameter family of initial conditions spanning a range of physically plausible initial eccentricities, deviations from the resonant period ratio, and libration amplitudes around the resonance centers.

\section{Simulations} \label{simulations}

We ran the above procedure to generate one thousand sets of initial conditions, 368 of which successfully captured at the observed period ratio commensurabilities. We drew $K$ randomly from a log-uniform distribution over the range $[10,10^3]$. This yielded eccentricities for the planets in strong first-order MMRs $\approx 0.01-0.1$, and $\sim 10^{-3}$ for the other two planets $b$ and $c$, and bracketed the range that matches observed systems \citep[e.g.,][]{Lee02} and hydrodynamical simulations \citep[e.g.,][]{Kley04}. We independently drew each value of $\eta$ from a log-uniform distribution between $10^{-3}$ (the system's minimum relevant scale for deviations from equilibrium set by gravitational kicks at conjunctions) and unity, which should initialize resonances close to their respective chaotic separatrices.

All integrations were performed using the {\sc \tt WHFAST} integrator \citep{ReinTamayo15} in the open-source {\sc \tt REBOUND} N-body package \citep{Rein12}.
We adopted a timestep of 7\% of the innermost planet's orbital period. The migration and eccentricity damping forces used during the generation of initial conditions as described in Sec.\:\ref{genic} were applied using the \texttt{modify\_orbits\_forces} routine in the REBOUNDx\footnote{\url{https://github.com/dtamayo/reboundx}} library.

\begin{figure}
 \centering \resizebox{\columnwidth}{!}{\includegraphics{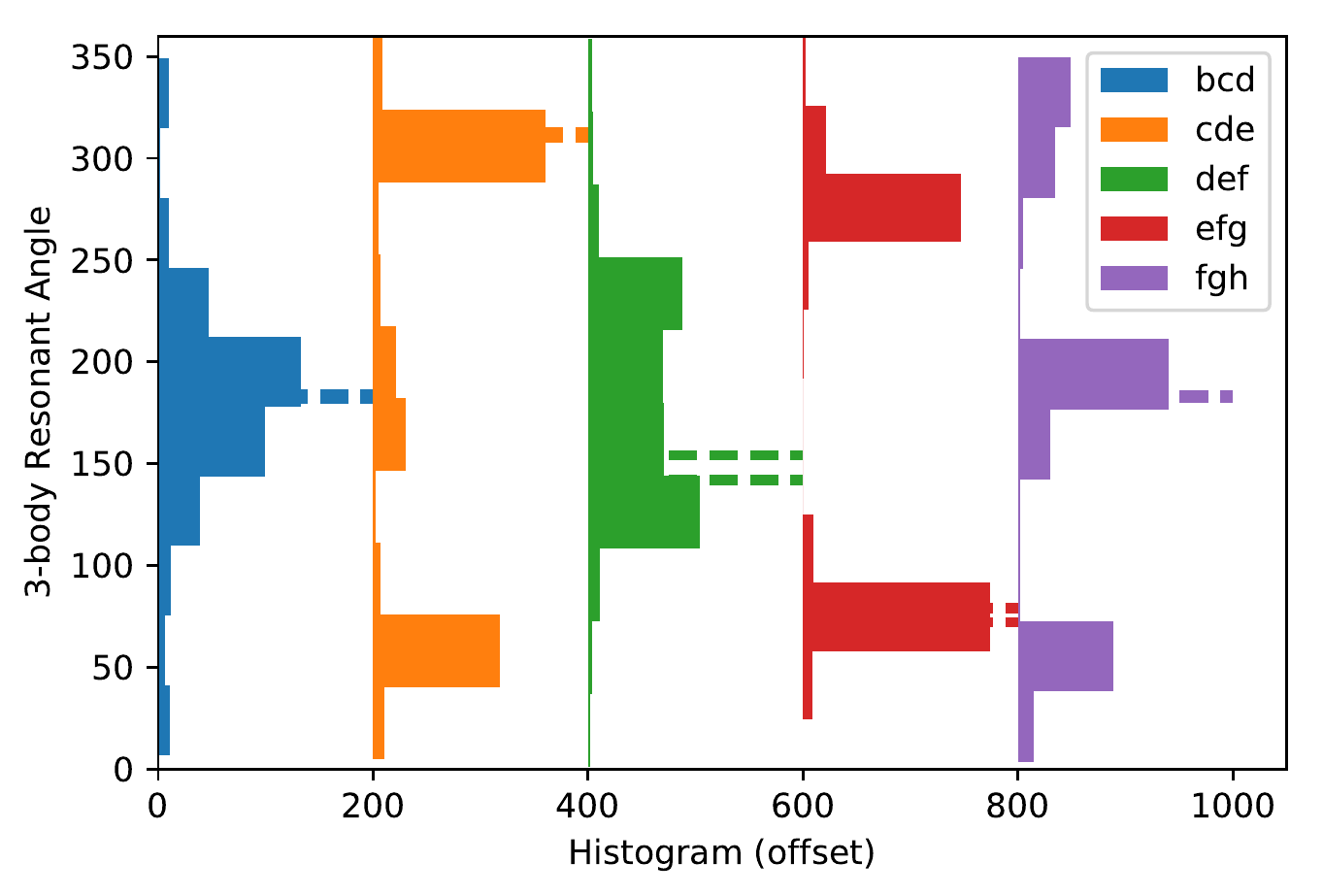}}
 \caption{ Distribution of the initial three-body resonance angles across our suite of simulations. The histograms have been offset for clarity. The dashed lines correspond to the observed range (see text).
\label{librationcenters}}
\end{figure}

The time baseline of observations of TRAPPIST-1 is not long enough yet for a determination of the centers and libration amplitudes of the current three-body resonance angles \citep{Luger17}. We can nevertheless check for a qualitative match between our suite of initial conditions and the observed system. In Fig.\:\ref{librationcenters}, we plot histograms of the initial three-body resonance angles in our simulations, with dashed lines depicting the range over which the corresponding angles have been observed to vary by \cite{Luger17}. Because the full range of variation in the observed system is not yet known, we do not try to restrict the systems we consider.

All 368 initial conditions were run for 5 Myr ($\gtrsim 10^9$ orbits of the innermost planet, and an order of magnitude longer than the instability timescale reported by \citealt{Gillon17}), and systems were marked as unstable if any pair of planets came closer to one another than the innermost planet's Hill radius, or if any planet was scattered beyond $1$ AU. A random subsample of 21 initial conditions were additionally run to 50 Myr. 

The precession induced by post-Newtonian corrections, the star's oblateness, and the stellar and planetary tides are all negligible compared to the precession induced by the planets' Newtonian point-source gravitational interactions.
The migration induced by tides raised on the star is slow, and there is not enough angular momentum in the planetary spins to appreciably change their semimajor axes as they tidally lock.

The only important effect beyond point-source gravity should be eccentricity damping caused by the equilibrium tides raised by the star on the planets \citep{Gillon17}. However, to isolate the effect of initial conditions, and to assess whether the system's stability can constrain the system's tidal dissipation history, we considered conservative systems interacting only through point-source gravity. This additionally circumvented subtle numerical artifacts that can arise when adding velocity-dependent forces to otherwise symplectic schemes for long-term integrations, and provides a stringent test of stability since eccentricity damping should help stabilize the system.

All integrations were saved in \texttt{SimulationArchive} format \citep{ReinTamayo17} and we make them publicly available. This makes it possible for other users to reproduce the simulations bit by bit on different machine architectures.

\section{Results}

We begin by comparing the above suite of simulations to integrations where we drew initial conditions from the parameters reported in \cite{Gillon17}. We treated all parameters listed in their Table 1 as independent Gaussian random variables with standard deviations given by their errors, and integrated 200 such systems. 
\begin{figure}
 \centering \resizebox{\columnwidth}{!}{\includegraphics{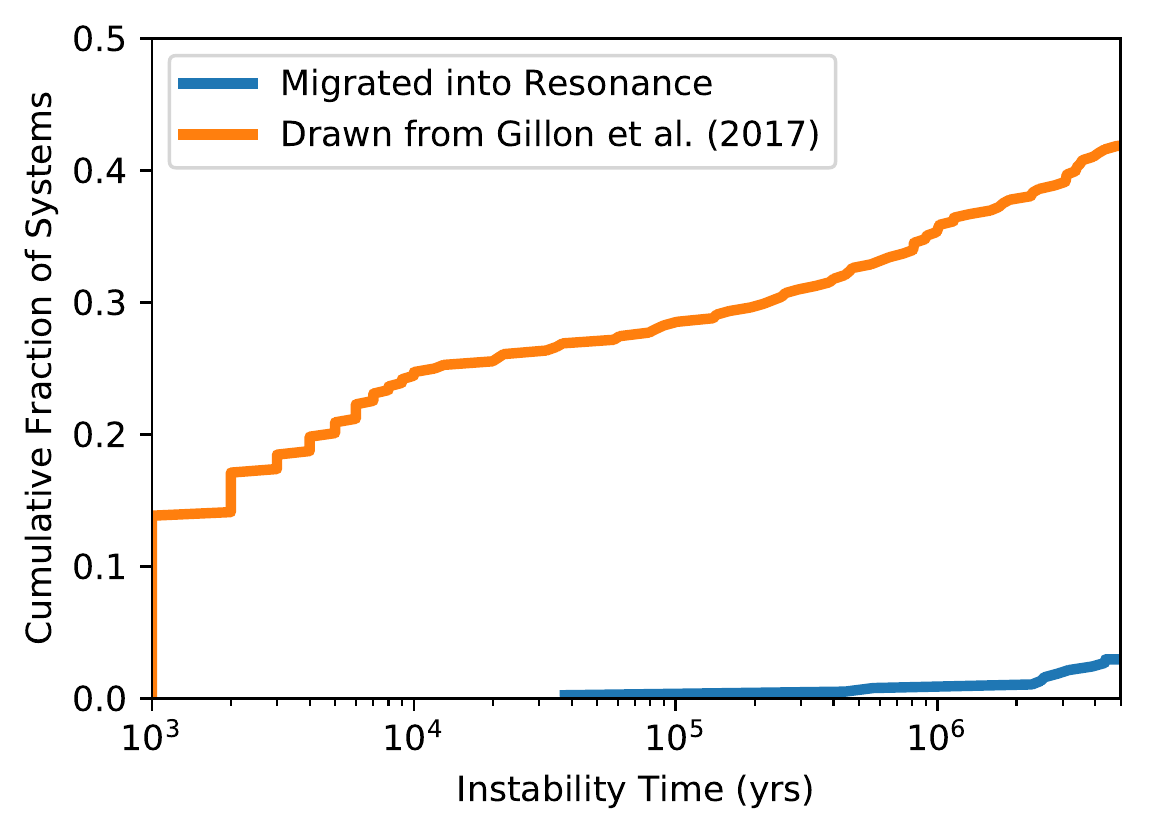}}
 \caption{ Cumulative distribution of instability times for a suite (orange) of initial conditions drawn from parameters reported by \cite{Gillon17}, and our set (blue) generated through disk migration.
\label{tinst}}
\end{figure}

As seen in Fig.\:\ref{tinst}, and found by \cite{Gillon17}, many initial conditions drawn from their fits go unstable on short timescales, with $\approx 42\%$ destabilizing within 5 Myr (orange curve). By contrast, only $\approx 3\%$ of our configurations initialized through disk migration were unstable. Of the 21 disk-migration systems additionally run to 50 Myr, only four ($\approx 19\%$) went unstable. 

Thus, most initial conditions migrated into resonance over a wide range of $K$ and $\eta$ were stable over at least 50 Myr, two orders of magnitude longer than the instability timescale reported by \cite{Gillon17} for initial conditions drawn from their posteriors. This highlights the value of imposing long-term stability to constrain orbital parameters in tightly packed systems like TRAPPIST-1, and provides strong evidence for slow migration adiabatically moving the system into an equilibrium resonant configuration.

As one can see from the definitions in Sec.\:\ref{genic}, two-body resonance angles involve one of the two planets' longitudes of pericenter, an angle which is difficult to constrain observationally for low-eccentricity orbits. By contrast, three-body resonance angles involve only the mean longitudes, which are precisely measured by transit observations. For our analysis, we therefore concentrate on the more easily observable three-body resonance angles, which, being linear combinations of two-body resonance angles, indirectly track the two-body MMRs.

To reduce each system's dimensionality for visualization, we calculated the full libration amplitude (difference between the maximum and minimum value) over a 0.5 Myr window for each three-body resonance angle and we hereafter refer to the maximum libration amplitude across the three-body resonances as the {\it system's libration amplitude}. We calculated each system's initial and final libration amplitude by performing the above procedure over the first and last 0.5 Myrs of the integration. The two libration amplitudes therefore lie on the interval $[0,360^\circ]$, where values $\approx 360^\circ$ mean that at least one three-body resonance angle is circulating.

\begin{figure}
 \centering \resizebox{\columnwidth}{!}{\includegraphics{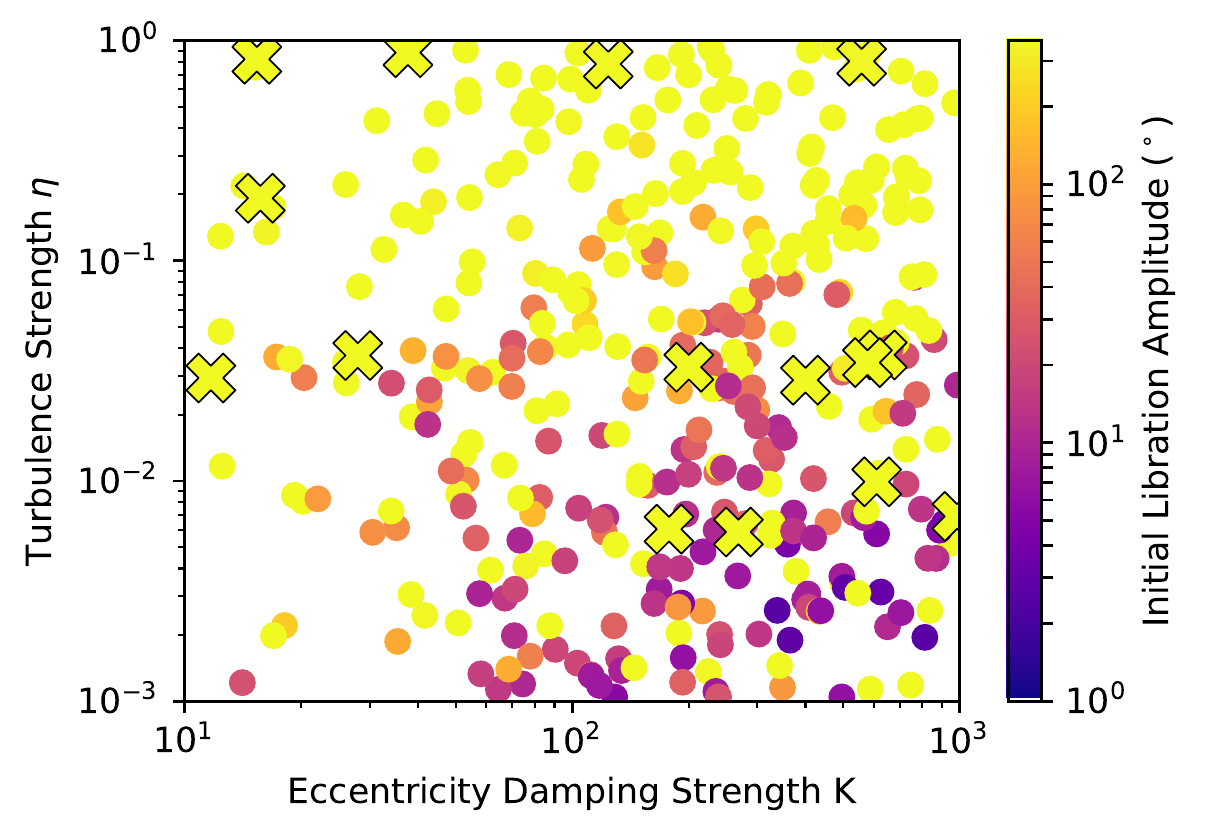}}
 \caption{ Initial libration amplitudes on the grid of resonant chains derived from disk migration. Initial eccentricities are lower to the right, and libration amplitudes increase upward. Unstable systems are marked with crosses, all of which initially circulate (libration amplitude $\approx 360^{\circ}$).
\label{ic}}
\end{figure}

Figure \ref{ic} shows each system's initial libration amplitude in the $K-\eta$ space we used to generate our initial conditions. Systems tend to be knocked out of resonance (libration amplitude $\approx 360^\circ$, yellow color) when the disk's eccentricity damping is weak (low $K$) and/or when the turbulence is strong (high $\eta$).
Initial conditions that went unstable are marked in Fig.\:\ref{ic} as crosses. We note that all such systems initially had at least one circulating three-body resonance angle, so we suspect that systems with {\it small} libration amplitudes (corresponding to low $\eta$) are stable on timescales much longer than 50 Myr, even without tidal dissipation.

If the above is true, one would expect that low-libration-amplitude systems should be nearly quasi-periodic and maintain their original libration amplitudes on long timescales. By contrast, systems initialized with progressively larger libration amplitudes should encounter chaotic regions around resonance separatrices that will move resonance angles in and out of circulation, and drive up eccentricities as the system diffuses through action space \citep[e.g.,][]{Murray99}.

\begin{figure}
 \centering \resizebox{\columnwidth}{!}{\includegraphics{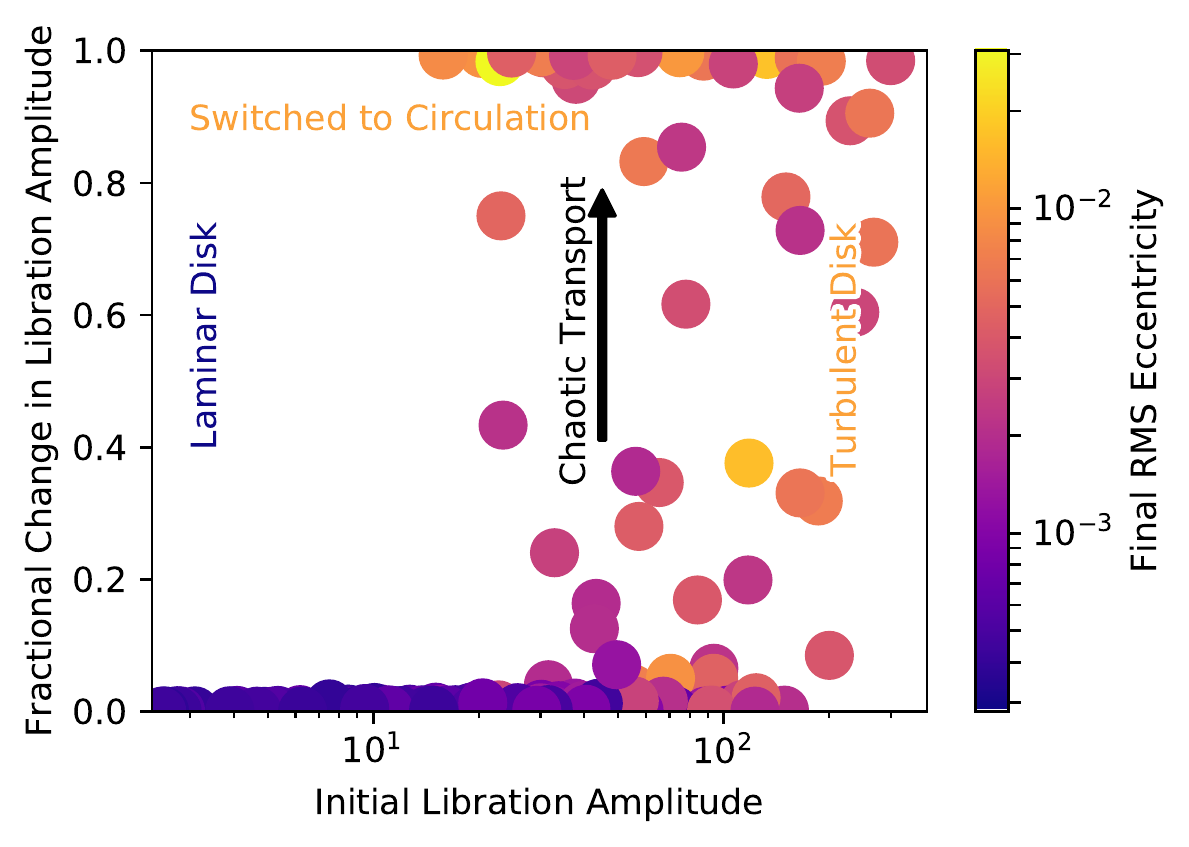}}
 \caption{ Fractional change in libration amplitude vs initial libration amplitude. Systems starting with low libration amplitude remain in the same configuration at small eccentricity. At larger libration amplitude, systems diffuse toward higher libration amplitudes and eccentricities. See text for details.
\label{evolution}}
\end{figure}

We see such behavior in Fig.\:\ref{evolution}, which considers all systems with initial libration amplitude $< 300^\circ$. It plots the fractional change in the libration amplitude over the span of the integration (zero represents no change, while unity corresponds to systems that switched to circulation) vs the initial libration amplitude. The color scale encodes each system's maximum RMS eccentricity across the seven planets over the last 0.5 Myr of the integration.

We see that systems that begin at low libration amplitude remain there throughout the integration, and maintain low RMS eccentricities. Beyond $\sim 10^\circ$ libration amplitude, the libration amplitude grows throughout the simulation, and this is associated with higher RMS eccentricities. The fact that the libration amplitudes of systems with initial values $<10^\circ$ do not evolve significantly over the integration suggests that they are likely to be stable over much longer timescales than we were computationally able to probe.

\section{Conclusion}
The TRAPPIST-1 system is remarkably perched near a chain of integer period ratios. The rich phase space structure created by these nearby MMRs puts strong constraints on the system's initial conditions if it is to remain stable for its age. In fact, most initial conditions drawn from the system's observationally inferred parameters go unstable on short timescales \citep{Gillon17}. This reflects the difficulty in matching the eccentricities and orbital alignments of any stable resonant islands given the observational uncertainties.

By contrast, we find that even without any tidal dissipation, most systems initialized through disk migration with a wide range of parameters are stable on timescales of at least 50 Myr, two orders of magnitude longer than the typical stability times found by \citep{Gillon17}. We expect that the inclusion of eccentricity damping through equilibrium tides raised on the planets should maintain most of our initial conditions stable over the system's age.

The observed orbital period ratios and the stability that we have found for systems initialized through disk migration provide strong evidence for a history of slow, convergent migration in the TRAPPIST-1 system. While in this manuscript we assumed that this migration occurred in the protoplanetary disk, we note that other migration mechanisms are also possible, e.g., interactions with a planetesimal disk.

The observationally inferred parameters should improve as transit timing variations are monitored over a longer baseline, but this study suggests that imposing long-term stability will also strongly constrain and help characterize the system.\footnote{See also \cite{Quarles17}, which appeared on arXiv at the same time this manuscript was submitted.}

The computational cost of long-term N-body simulations will make such attempts challenging; \cite{Tamayo16} recently demonstrated that machine-learning algorithms can be trained to predict the stability of three-planet systems quickly and with high fidelity; it would be valuable to further develop such methods for systems as dynamically complex and long-lived as TRAPPIST-1.

Once the three-body resonance angles identified by \cite{Luger17} complete a full cycle, an obvious extension of this work would be to incorporate tidal effects and to attempt to match the observed centers and libration amplitudes of any librating angles, starting from physically motivated initial conditions like the ones presented here.  We have made all simulations and source code public, and by using machine-independent algorithms \citep{ReinTamayo17} make it straightforward for others to reproduce and build upon this work.

This would not only inform formation conditions in the disk (the parameters in Fig.\:\ref{ic}), but also the tidal history of the system. While our results suggest that dissipation is not necessary for {\it stability}, tides will modify the three-body resonance angles, and dissipation is required to spread the resonant chain \citep{Batygin12, Lithwick12} and push the period ratios $\approx 1\%$ wide of their respective nominal resonant values, as currently observed \citep{Gillon17}. Thus, matching the period ratios and three-body resonance angles should set important limits on the planetary tidal parameters and their associated tidal heating rates.

The simulation archives generated for this work are openly available at \dataset[https://doi.org/10.5281/zenodo.496153]{https://doi.org/10.5281/zenodo.496153}. The scripts used to run the simulations, process and visualize the data, and generate the figures in this manuscript are available at \url{https://github.com/dtamayo/trappist}.

\bigskip

We would first like to thank John Dubinski and Claire Yu, whose technical and computing support were essential to this project. We would also like to thank Alan Jackson, Kristen Menou and Yanqin Wu for insightful discussions. D.T. is grateful for support from the Jeffrey L. Bishop Fellowship. H.R. was supported by NSERC Discovery Grant RGPIN-2014-04553. C.P. is grateful for support from the Gruber Foundation Fellowship. This research was made possible by the computing resources at the Canadian Institute for Theoretical Astrophysics and the open-source projects listed below. \software{\texttt{Jupyter} \citep{jupyter}, \texttt{iPython} \citep{ipython}, \texttt{matplotlib} \citep{matplotlib, matplotlib2}}.

\end{document}